\documentclass[final,1p,times]{elsarticle}

\usepackage{amssymb}

\usepackage[hidelinks]{hyperref}

\usepackage{todonotes}

\usepackage{listings}
\colorlet{punct}{red!60!black}
\definecolor{background}{HTML}{EEEEEE}
\definecolor{delim}{RGB}{20,105,176}
\colorlet{numb}{magenta!60!black}
\lstdefinelanguage{json}{
    basicstyle=\footnotesize\ttfamily\linespread{0.05},
    numbers=none,
    stepnumber=0.1,
    numbersep=pt,
    tabsize=1,
    showstringspaces=false,
    breaklines=true,
    columns=fullflexible,
    literate=
     *{0}{{{\color{numb}0}}}{1}
      {1}{{{\color{numb}1}}}{1}
      {2}{{{\color{numb}2}}}{1}
      {3}{{{\color{numb}3}}}{1}
      {4}{{{\color{numb}4}}}{1}
      {5}{{{\color{numb}5}}}{1}
      {6}{{{\color{numb}6}}}{1}
      {7}{{{\color{numb}7}}}{1}
      {8}{{{\color{numb}8}}}{1}
      {9}{{{\color{numb}9}}}{1}
      {:}{{{\color{punct}{:}}}}{1}
      {,}{{{\color{punct}{,}}}}{1}
      {\{}{{{\color{delim}{\{}}}}{1}
      {\}}{{{\color{delim}{\}}}}}{1}
      {[}{{{\color{delim}{[}}}}{1}
      {]}{{{\color{delim}{]}}}}{1},
}

\usepackage[flushleft]{threeparttable}

\newcommand\notype[1]{\unskip}

\usepackage[addedmarkup=uuline]{changes}
\definechangesauthor[color=red]{deleted}

\journal{Advanced Engineering Informatics}

\begin{document}

\begin{frontmatter}



\title{Applying Digital Twins for the Management of Information in Turnaround Event Operations in Commercial Airports}


\author[labelupm]{Javier Conde\corref{cor1}}
\ead{javier.conde.diaz@upm.es}
\author[labelupm]{Andres Munoz-Arcentales}
\author[labelkth]{Mario Romero}
\author[labelferro]{Javier Rojo}
\author[labelupm]{Joaquín Salvachúa}
\author[labelupm]{Gabriel Huecas}
\author[labelupm]{Álvaro Alonso}

\cortext[cor1]{Corresponding authors.}
\fntext[]{Publication accepted and published in: https://doi.org/10.1016/j.aei.2022.101723}

\affiliation[labelupm]{organization={\unexpanded{Departamento de Ingeniería de Sistemas Telemáticos, Escuela Técnica Superior de Ingenieros de Telecomunicación, Universidad Politécnica de Madrid}},
            city={Madrid},
            country={Spain}}
            
\affiliation[labelkth]{organization={\unexpanded{Department of Computational Science and Technology, School of Electrical Engineering and Computer Science, KTH Royal Institute of Technology}},
            city={Stockholm},
            country={Sweden}}
            
\affiliation[labelferro]{organization={\unexpanded{Ferrovial}},
            city={Madrid},
            country={Spain}}

\begin{abstract}
The aerospace sector is one of the many sectors in which large amounts of data are generated. Thanks to the evolution of technology, these data can be exploited in several ways to improve the operation and management of industrial processes. However, to achieve this goal, it is necessary to define architectures and data models that allow to manage and homogenise the heterogeneous data collected. In this paper, we present an Airport Digital Twin Reference Conceptualisation's and data model based on FIWARE Generic Enablers and the \added{Next Generation Service Interfaces-Linked Data} standard. Concretely, we particularise the Airport Digital Twin to improve the efficiency of flight turnaround events. The architecture proposed is validated in the Aberdeen International Airport with the aim of reducing delays in commercial flights. The implementation includes an application that shows the real state of the airport, combining \added{two-dimensional} and \added{three-dimensional} virtual reality representations of the stands, and a mobile application that helps ground operators to schedule departure and arrival flights.
\end{abstract}




\end{frontmatter}



\section{Introduction}
\label{sec:intro}

The communication between the virtual and physical world constitutes the basis for industry 4.0, being the cyber-physical systems (CPS), digital twins (DTs), and the Internet of Things (IoT) their key elements \cite{pr10040744, 9626349, app10186519}. Although the concept of DT has emerged strongly in recent years, its origin comes from Michael Grieves who presented it in 2003 \cite{grieves2014digital}, and the first official definition of DT was provided by the \added{National Aeronautics and Space Administration} (NASA) \cite{nasa}. Grieves and John Vickers coined the term digital twin to refer to a `virtual digital equivalent to a physical product'. Since its origin, due to the evolution of technology, new concepts related to DT and definitions of DT have emerged and, as a result, \added{today there is no total consensus \cite{iceis22}}. The research conducted by Sjarov et al. \cite{9212089} collects different terms present in the literature as synonyms for DT or closely related to them (e.g., Digital Shadow, Digital Surrogate, Digital Triplet, Product Avatar, Virtual Twin). From all concepts found in the literature, Kritzinger et al. \cite{KRITZINGER20181016} define a DT as the digital counterpart of a physical object. Moreover, they classify a DT into three categories depending on the level of integration between the physical and the digital object. They refer to Digital Model when the flow of data is manual from the physical to the digital object, and vice versa. It is called Digital Shadow when the flow of data is automatic from the physical object to the digital one, but manual in the other direction. In other words, some change in the state of the physical object may produce an automatic change in the state of the digital one, but not in the reverse direction. Lastly, it is known as Digital Twin when the flow of data is automatic in both directions, i.e., some change in the state of the physical object may produce a change in the state of the digital one and vice versa.  

In the work of Grieves~\cite{grieves2014digital}, he states that any DT is made up of three parts: (1) the Real Space containing the physical products; (2) the Virtual Space containing the virtual products; and (3) the connections between the physical and virtual products. Four years later, Tao et al. \cite{TAO2018169} expanded this definition with two more dimensions: (4) the DT data manager which controls the data of the physical and the virtual counterpart, and (5) the external services which interact with the DT, such as client applications or data suppliers. The literature review conducted by Jones et al. \cite{JONES202036} deepens in the definition of DTs proposing new characteristics. According to Jones, a DT is composed of physical entities that perform their activities through predefined physical processes in a physical environment. For each physical entity, there are a set of virtual entities operating in a virtual environment that constitute their representation in the virtual world. In contrast to the physical environment, the virtual one may receive data from external sources such as \added{application programming interfaces} (APIs) or databases. Physical and virtual entities are defined by their state. In other words, entities are defined by a set of parameters that characterise them. These parameters can be static (e.g., geometry) or dynamic (e.g., temperature) \cite{SCHROEDER201612}. Jones states that a key element of any DT is the synchronisation between physical and virtual entities. This process is known as twinning, and the frequency of occurrence is determined by the twinning rate. The twinning process is composed of two stages. Firstly, an event happens in the physical/virtual world and it is captured (metrology phase); secondly, the change is transmitted to the virtual/physical entities affected (realisation phase). The twinning process may occur in both directions, from the physical entities to the virtual ones (physical-to-virtual connection), or from the virtual entities to the physical ones (virtual-to-physical connection). 

Besides the definition of common characteristics among DTs, authors also mention barriers for implementing them. One of the most mentioned issues is the problem of working simultaneously with the physical and virtual world because it makes it harder to manage dynamic information which requires real-time guarantees \cite{LIU2021346, UHLEMANN2017335}. Other challenges mentioned are the difficulty of integrating heterogeneous sources of data, managing a large amount of data in distributed environments, validating the correct operation of the DT, standardising the communication between all pieces that interact with the DT, guaranteeing the security of the system, and managing the access of different actors \cite{9103025, pr10040744}.

Since its origin, DTs have been applied in the military and aeronautics industry \cite{QI20213}, however, over the years, \added{new use cases have emerged in multiples fields as manufacturing, construction, transportation or healthcare \cite{9781003017547}. In the last decade, DTs have emerged in the manufacturing sector thanks to IoT sensorisation \cite{2021.2022762}. The most widespread use cases are the improvement of the efficiency of physical machines and the optimisation of processes \cite{2021.2022762}. The DT captures real-time events from the factory, processes them, monitors the operation, and alerts of potential faults \cite{FRIEDERICH2022103586}. Scime et al. \cite{SCIME202228} propose the division of complex tasks into single operations that can be combined through digital threads. The simplification into simple operations facilitates the monitoring and scaling of DTs.} \added{In the architecture, engineering, and construction (AEC) sector DTs can be applied in all phases from design to construction \cite{SASBE-08-2021-0148}. Alizadehsalehi and Yitmen \cite{SASBE-01-2021-0016} propose the use of DT together with extended reality (XR), reality capturing technologies, and Building Information Modelling (BIM) to monitor the construction process. In addition, other works propose the use of Deep Learning to improve the decision-making process, progressing toward a Cognitive Digital Twin (CDT) \cite{app11094276}. CDTs are an evolution of DTs, with cognitive capabilities, able to perform tasks in a similar way as a human would do \cite{00207543.2021.2014591}. In the field of healthcare, DTs have been used to manage healthcare facilities by predicting failures and implementing preventive maintenance \cite{9780784483893.046}. However, Sony and Li \cite{9780784483961.120} point out that the lack of coordination between hospital departments is the main barrier to applying DTs in the healthcare system. DTs have also gained importance in other fields such as transportation or smart agriculture \cite{JONES202036}.}

This manuscript reports the experience of implementing a Digital Twin at Aberdeen International Airport used for monitoring and managing the turnaround events of an aircraft, i.e., all the events that happen in an aircraft since its arrival to the airport until its departure. The implementation of the DT has followed the guidelines proposed in a previous research that presents the FIWARE Initiative as an enabling environment for developing DTs and modelling their data \cite{9346030}. FIWARE~\footnote{\url{https://www.fiware.org/}} is an Open Source framework made up of software components that eases the implementation of smart solutions, including among them, DTs.  

The article is structured as follows. In the next section, related work on DTs and DTs applied to the aerospace sector is reviewed. In Section~\ref{sec:architecture}, we present the airport reference architecture and the airport data model based on the FIWARE ecosystem tools; we also explain how the information is processed through the whole architecture and the barriers found when implementing the DT. Section~\ref{sec:visualization} explains some services that interact with the DT, concretely the client applications that help the airport workers to manage turnaround operations. In Section~\ref{sec:extension}, we discuss the extension of our proposals to other airports and its possible challenges. Finally, in Section \ref{sec:conclusions}, conclusions and future research are proposed.


\section{Related work}
\label{sec:related}

Over the years, some studies have been carried out analysing the knowledge and technology required to guarantee the correct operation of DTs \cite{QI20213, 9103025}. Some researches are focused on the physical world. These works study the measuring equipment used for collecting data and the requirements for building physics models that represent the real world through mathematical equations \cite{QI20213}. Regarding the virtual entity, there are studies in the fields of monitoring, visualisation, and validation tools \cite{4046739}. Other works are specialised in data management, with topics such as storage of large amounts of data, data processing, or data formatting \cite{CIAVOTTA2017931, 8258937}. Some researchers study the interconnection between the elements that compound a DT, delving into Internet technologies, communication protocols, IoT, etc. \cite{app10186519}. 

In addition to the already mentioned relationship between IoT and DTs \cite{app10186519}, there are studies focused on the inclusion of Artificial Intelligence (AI), Machine Learning (ML), and Deep Learning (DL) within DTs \cite{9103025, MIN2019502}, and the use of Big Data techniques for processing large amounts of data \cite{9359733}. The combination of these technologies adds a new layer of integration known as Intelligent Digital Twin \cite{AshtariTalkhestani}. Intelligent Digital Twins propose using historical data collected from the physical entity to build AI models and apply them in the virtual world. The outcomes can be used for detecting, predicting, optimising, and making decisions about the physical entity \cite{9359733}.

\subsection{\added{Architectures for developing DTs}}
When implementing a DT, most authors propose their own solutions designed by themselves. However, almost all of them have in common the elements mentioned by Grieves \cite{grieves2014digital}, Tao et al. \cite{TAO2018169} and Jones et al. \cite{JONES202036}. \added{They are layered architectures focused on the flow of data process \cite{9770073}}. The first step is data acquiring from external sources (e.g., physical devices, databases, APIs), followed by data modelling, continuing with data processing, and ending with data consuming by different clients (e.g., web application, \added{three-dimensional} [3D] application, physical devices) \cite{9770073}. Some proposals analyse the communication between the physical and virtual world and data storage instead of data processing and client applications \cite{9103025, 03003}. The work of Redelinghuys et al. \cite{03003} proposes an architecture of six layers where data from physical devices are stored in local repositories and in the cloud to be consumed by simulation systems. Other architectures are valid for scalable distributed environments. In this sense, Ciavotta et al. \cite{CIAVOTTA2017931} present MAYA, a platform for developing scalable DTs in smart factories. Picone et al. \cite{PICONE2021100661} present an architecture based on the White Label Digital Twin (WLDT) Java library and an orchestrator called WLDT Engine. In this architecture, the orchestrator controls a set of workers that operate the DT through a software processing pipeline. Other authors explain architectures for building intelligent DTs based on Artificial Intelligence, Machine Learning, or Deep Learning \cite{9359733, AshtariTalkhestani}. These works not only mention AI algorithms, but also the Big Data infrastructure required to manage large amounts of data. The proposal of Talkhestani et al. includes the collaboration among different DTs as a mechanism for improving the results. With a similar perspective, Redelinghuys et al. \cite{10079783} expanded their reference architecture to enable the aggregation of DTs. Moreover, some architectures focus on specific use cases, for example, Li et al. \cite{LI2021108223} explain how to use blockchain to ensure secure knowledge sharing among all the actors of a DT, and Yun et al. \cite{7993933} present a platform called uDit \added{(universal Digital Twin platform)} to include Quality of Service (QoS) policies in DTs.

\subsection{\added{Platforms for developing DTs}}
Besides reference architectures, there are commercial platforms for building DTs providing a higher level of abstraction. Most of them are platform as a service solutions (PaaS) offered by cloud providers (e.g., Azure Digital Twins~\footnote{\url{https://docs.microsoft.com/en-us/azure/digital-twins}}, Vertex~\footnote{\url{https://aws.amazon.com/es/iot/solutions/VertexDigitalTwin}}, \added{International Business Machines Corporation} [IBM] Digital Twin Exchange~\footnote{\url{https://www.ibm.com/products/digital-twin-exchange}}). These platforms are composed by a set of tools that ease the integration of IoT devices, data modelling, data storage, security, etc. However, as a consequence of their PaaS model, they do not allow to control the whole architecture because they hide the complexity of their tools offering them as black boxes. With the same aim, there are open projects for supporting DTs. This is the case of iTwin Platform~\footnote{\url{https://developer.bentley.com/}}, Eclipse Ditto~\footnote{\url{https://www.eclipse.org/ditto/}} or FIWARE~\footnote{\url{https://www.fiware.org/}}. The iTwin platform is mainly oriented to the visualisation counterpart of a DT, integrating \added{two-dimensional} (2D) and \added{three-dimensional} (3D) models in client applications. Both Eclipse Ditto and FIWARE offer a set of software components called, respectively, Ditto Services and FIWARE Generic Enablers (GEs). These software components ease the development and management of context information in DTs. These software pieces are highly configurable and provide the capability to implement the functionality of a DT by (1) capturing data from the real world, mainly through IoT sensors, and transmitting them to the DT, and vice versa; (2) homogenising the information against predefined data models; (3) managing the current state of the DT; (4) processing the data; and (5) supplying the data to client applications. In the literature, there are researches that have implemented DTs using Eclipse Ditto \cite{9119497} and FIWARE \cite{0951192X}.  

\subsection{\added{Data modelling in DTs}}
The centrepiece of a DT is the data modelling \cite{Boschert2016}. Data models define how the real world is represented in the virtual environment. According to Liu et al. \cite{LIU2021346}, there are two types of models, physics models and semantic models. On the one hand, physics models represent the behaviour of a system through physical equations. On the other hand, semantic models capture the current status of physical objects and manage them in their virtual counterparts, using, for example, AI techniques. Semantic models do not require complex equations, instead they use historical data to generate knowledge. The use of physical or semantic models is not exclusive, they can be combined. Chakraborty and Adhikari \cite{CHAKRABORTY2021106410} propose a hybrid scheme which combines machine learning algorithms with physics models. However, the integration of interdisciplinary models brings challenges, including their storage, exchange with other systems, and processing \cite{AshtariTalkhestani}. Consequently, it is important to define standards and data formats to represent DT models. The research conducted by Jacoby and Usländer \cite{app10186519} analyses the most extended standards used for modelling DTs. Concretely, they propose the Asset Administration Shell (AAS), the Next Generation Service Interfaces-Linked Data (NGSI-LD), the Digital Twin Definition Language (DTDL), the Open Data Protocol (OData), the SensorThings API (STA), and the Web of Things (WoT). Among the standards mentioned, the most extended model languages and data serialisation formats are \added{Automation Markup Language} (AutomationML), \added{Unified Modeling Language} (UML), \added{Systems Modeling Language} (SysML), \added{eXtensible Markup Language} XML, \added{JavaScript Object Notation} (JSON), \added{JavaScript Object Notation-Linked Data} (JSON-LD), and \added{Resource Description Framework} (RDF) \cite{app10186519}.

\subsection{\added{FIWARE for developing DTs}}
As stated in the introduction, our research is based on a previous work that proposes a reference architecture for building DTs based on the FIWARE GEs, and the FIWARE Smart Data Models, completed with a theoretical example of a Parking Digital Twin \cite{9346030}. Our investigation validates the proposal of the previous research within a real use case of a Digital Twin Airport. The FIWARE foundation defines a DT as `an entity which digitally represents a real-world physical asset' \cite{ffdts}. In FIWARE, the state of virtual entities is called context and it is managed by the Context Broker GE. Context can be accessed synchronously or asynchronously using the Next Generation Service Interfaces for Linked Data (NGSI-LD) implemented by the Context Broker. Synchronous access is established through \added{Hypertext Transfer Protocol} (HTTP) petitions originated from the client. Asynchronous communication consists of a system of notifications based on the publish-subscribe pattern implemented by the Context Broker GE \cite{s21217095}. 

In FIWARE, the twinning process from the physical to the virtual world starts with an IoT sensor capturing a change of a physical entity (metrology phase), and it finishes with the update of the context information (realisation phase). On the other hand, the twinning between the virtual and the physical world starts with an update of a virtual entity (metrology phase) and it finishes with an IoT actuator updating the state of the corresponding physical entities (realisation phase). For the twinning processes, the FIWARE catalogue offers the IoT Agent GEs, i.e., software components located between the IoT devices (or IoT gateways) and the Context Broker. IoT Agents are modules that translate the NGSI-LD petitions into the communication protocols of the IoT devices (e.g., \added{Lightweight machine to machine} [LWM2M], UltraLight) and vice versa. 

As mentioned above, one of the challenges of DT is acquiring data from external sources and adapting them to the DT data model. In FIWARE this challenge could be solved using the Draco GE. Draco is a dataflow management system based on Apache NiFi that is highly scalable and suitable for data transforming operations and data routing through a set of highly configurable processors and controllers. Another cornerstone when implementing DT is data modelling, since it allows to integrate external sources of data in a uniform, consistent, and standard way. In this sense, the adoption of standard data models widely spread in the industry eases the integration of the DT with other external systems, being the DT the data provider for those systems. The FIWARE Smart Data Models Initiative~\footnote{\url{https://smartdatamodels.org/}} provides the needed tools for both topics: (1) defining a common schema that enables the integration of external data sources, and (2) using standard data models adopted by the industry.

\subsection{\added{DTs in the aerospace sector}}
Our research presents a real case of a DT applied in the aerospace sector, where DTs have been used since the origin of the DT concept \cite{QI20213} and in many different processes, including manufacturing, operation, and maintenance \cite{PHANDEN2021174}. The work of Aydemir et al. \cite{AydemirHakan_DigitalTwin_Paradigm} proposes the use of Big Data and DTs in aircrafts, where up to 300.000 parameters are monitored on a regular flight. In the aerospace industry, DTs have been used in manufacturing machines to avoid causing damage to aircraft parts or making defective pieces \cite{LADJ2021168}. There are research studies that propose DTs for the maintenance and failure detection of different aircraft parts, such as engines \cite{lacaille_specific}, the fuel system, or the environmental control system \cite{cordelia_9319676}; for monitoring the performance of the rotor of a helicopter \cite{GUIVARCH2019133}; for predicting damage in the structure of the aircraft \cite{majumdar_10.2514/6.2013-1577, teugel_reengineering}, etc. DTs have also been applied in flight operations. For example, the work conducted by Zohdi \cite{ZOHDI2021113446} presents a framework for managing the flight route of firefighting aircrafts to optimise the spread of fire retardants; and the research conducted by Wong et al. \cite{wong_2020.1775299} applies DTs to improve air cargo loading operations for air freight transport.

Regarding the turnaround process, Makhloof et al. \cite{makhloof_real} define it as the set of activities carried out at the airport to prepare an aircraft that has arrived from an inbound flight to the next outbound flight operated by the same aircraft. Wu and Caves \cite{wu_transportation_planning} showed that airline scheduling and ground service operations are the main factors that influence the turnaround process. With a similar approach, Postrino et al. \cite{POSTORINO2020429} detected that disruptions in ground staff yield delays in flights, concluding that an availability of 80\% of ground operators causes delays in 50\% of the flights. Moreover, the post-pandemic situation requires 20\% more time for turnaround operations \cite{SCHULTZ2020101886}. Turnaround is a complex process that involves multiple tasks and actors (e.g., airlines, ground handlers, airlines, airport operators \cite{OKWIR2017183}). In 2010, many airports did not have the turnaround process digitalised and the communication among different actors was established through telephone calls \cite{makhloof_real}. Today, airports apply decision making systems, as the \added{Airport Collaborative Decision Making} (A-CDM) standard for European airports \cite{acdm_eurocontrol}. However, according to Okwir et al. \cite{OKWIR2017183} these standards are not integrated in an efficient way. \added{When digitalising the turnaround process it is required to know the real-time position of aircrafts. Schultz et al. propose the Automatic Dependent Surveillance Broadcast (ADS-B) system for the location of aircrafts in combination with the A-CDM standard \cite{SCHULTZ2022102164}. Other works focus on image processing for building a real-time map of the airport. For example, the work of Thai et al. \cite{THAI2022103590} presents a ground aircraft recognition system based on image processing with Neural Networks. Finally, other studies analyse the emulation of the airport traffic control system in different scenarios, including emergencies and critical situations \cite{saif2022}.}


\section{Airport Architecture and Data Management}
\label{sec:architecture}

This Section presents the airport reference architecture and data model built up on the FIWARE ecosystem. The first subsection presents the architecture to obtain data from the identified sources, model, store them and make them available for being consumed by applications. The second subsection presents the data models used for the airport scenario using FIWARE Data Models initiative. Finally, the third subsection details how the ingested data are processed according to the defined data models. \added{The proposal is based on the reference architecture presented by Conde et al. \cite{9346030}. It will also analyse the differences and similarities with the parking use case of that study.}  

\subsection{Airport DT Architecture}

\begin{figure*}[]
    \centering
    \includegraphics[scale=0.15]{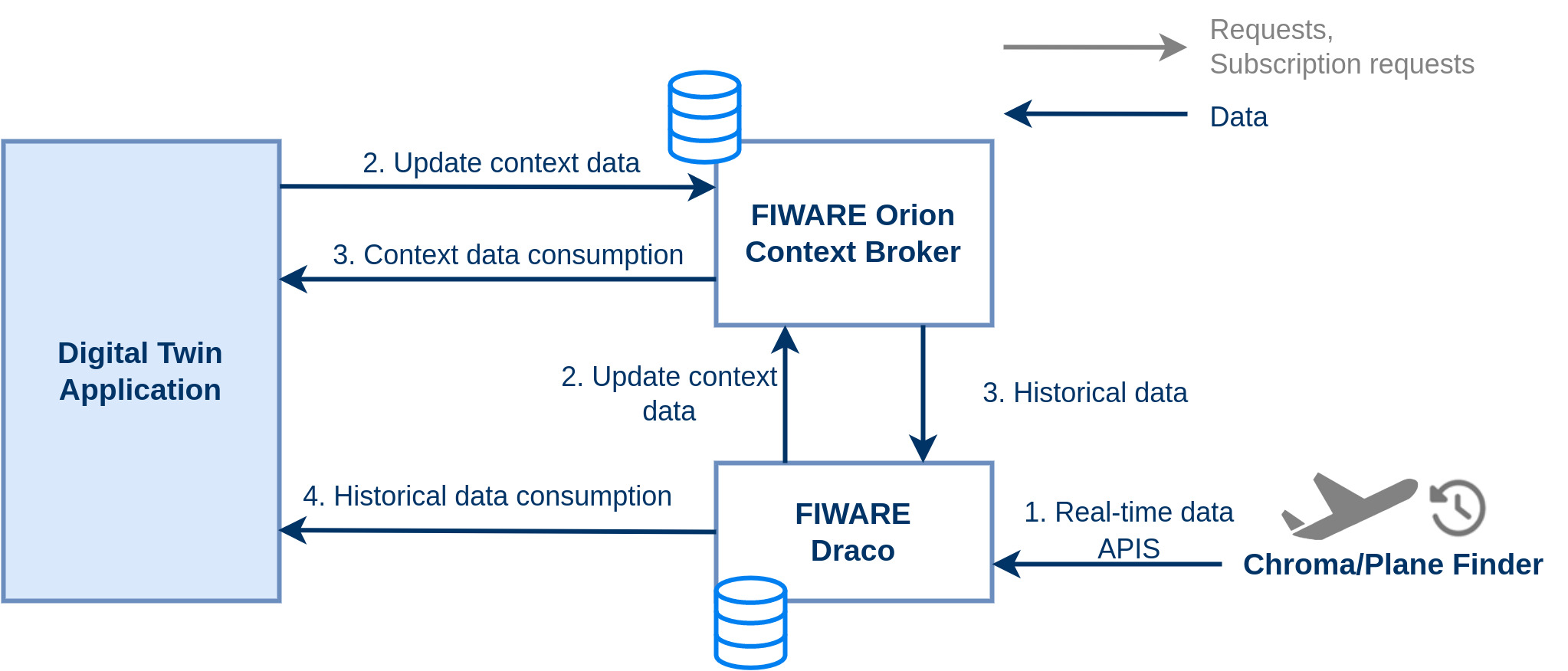}
    \caption{Airport DT architecture}
    \label{fig:general_architecture}
\end{figure*}

Before designing an architecture capable of providing all the services needed in an Airport Digital Twin, firstly we need to determine which data sources are presented in airports. After a deep analysis, we have identified that, in this field, we need to provide all the components needed not only for covering static data sources, but also for streaming. In this regard and as we stated in Section~\ref{sec:related}, our complete architecture relies on FIWARE GE’s and is shown in Figure~\ref{fig:general_architecture}. 

The central piece of the whole architecture, \added{as in the parking use case}, is the Context Broker, which is in charge of providing context management capabilities using NGSI as standard. \added{In contrast with the parking use case which was implemented with NGSIv2, the Airport DT extends the reference architecture using the NGSI linked data version (NGSI-LD)}. The implementation of the broker used in this proposal is known as Orion-LD. The Context Broker GE manages the entire life-cycle of context information including updates, queries, registrations, and subscriptions, for static data sources and for streaming. It manages context information through the implementation of a publish-subscribe system through an NGSI-LD interface. Users can create context elements, query and update them, and subscribe to changes in context information that they can receive as notifications. Other elements interact with the Context Broker through HTTP requests. The Context Broker offers the FIWARE NGSI-LD and NGSIv2 APIs, and an associated information model (entity, attribute, metadata) as the main interface for managing context data.

Thus, all the data represented as entities in the Context Broker need to be NGSI-LD compliant. In our proposal, we consider alternative data sources that can feed the Context Broker. Examples of these types of systems are \added{representational state transfer} (REST) APIs or database systems that need to be queried periodically, and systems like \added{Transmission Control Protocol} (TCP) or HTTP servers that send a continuous flow of data. These data, possibly coming from external applications, have to be converted into NGSI-LD and sent to the Context Broker. In this regard, our architecture relies on Draco GE. The suite of processors and controllers provided in this GE allows us to convert the incoming data into NGSI-LD entities and attributes. 

Additionally, in this architecture, the Context Broker provides a complete solution for managing the latest context data available, however, it does not maintain a history of the context data. In Section~\ref{sec:related}, we mentioned the importance of saving historical data for future processing or for building AI models. For that purpose, our reference implementation relies on the Draco GE, which receives each context update as a stream of data and injects them into multiple data storage systems for future processing.

Finally, the proposal is completed with two client applications. The first one is a web application that contains a real-time virtual representation of the airport through 2D and 3D models. The second one is a mobile application that helps operators to manage aircraft turnaround events.

The complete architecture of the Airport DT relies on FIWARE GEs and is shown in Figure~\ref{fig:general_architecture}. \added{It is based on the FIWARE reference architecture proposed for DTs \cite{9346030} but using NGSI-LD instead of NGSIv2. As the FIWARE ecosystem is agnostic to the use case, the GEs do not have to be modified for the Airport DT.}  

After the brief description of the components of the architecture, it is easy to present a workflow of the different stages where the data is transformed with the aim to build the DT. First, the data generated in streaming is collected and sent to Draco to be adapted to the NGSI-LD format. In the same way, the data coming from static data sources like databases or REST systems are also extracted and transformed with Draco. Once the received data are modelled, they are published in the Context Broker as entities. These entities are continuously updated depending on the changes received. Finally, these data will feed the DT applications via the subscription/notification mechanism provided by the Context Broker. On the one hand, the web application will represent the Aberdeen airport with information about flights, aircrafts, stands, etc. On the other hand, the mobile application will present notifications to airport operators, and will also allow them to register new events related to the turnaround process.

\subsection{Airport DT Data Model}

Besides building the DT following the reference architecture based on the FIWARE GEs proposed, it is also important to model the information. Our solution relies on the FIWARE Smart Data Models, due to their integration with other systems and the adoption of standards approved by the industry. \added{The parking DT represented the information with Smart Data Models that belong to the smart cities domain}~\footnote{\url{https://github.com/smart-data-models/SmartCities}}. \added{In this case, the Airport DT is built models from the aeronautics domain}~\footnote{\url{https://github.com/smart-data-models/SmartAeronautics}}.

The data models were built following the nomenclature used by the main international aviation organisations such as the International Civil Aviation Organization (ICAO~\footnote{\url{https://www.icao.int/}}); the International Air Transport Association (IATA~\footnote{\url{https://www.iata.org/}}); the European Organization for Safety of Air Navigation (Eurocontrol~\footnote{\url{https://www.eurocontrol.int/}}); etc.

\begin{figure*}
    \centering
    \includegraphics[scale=0.15]{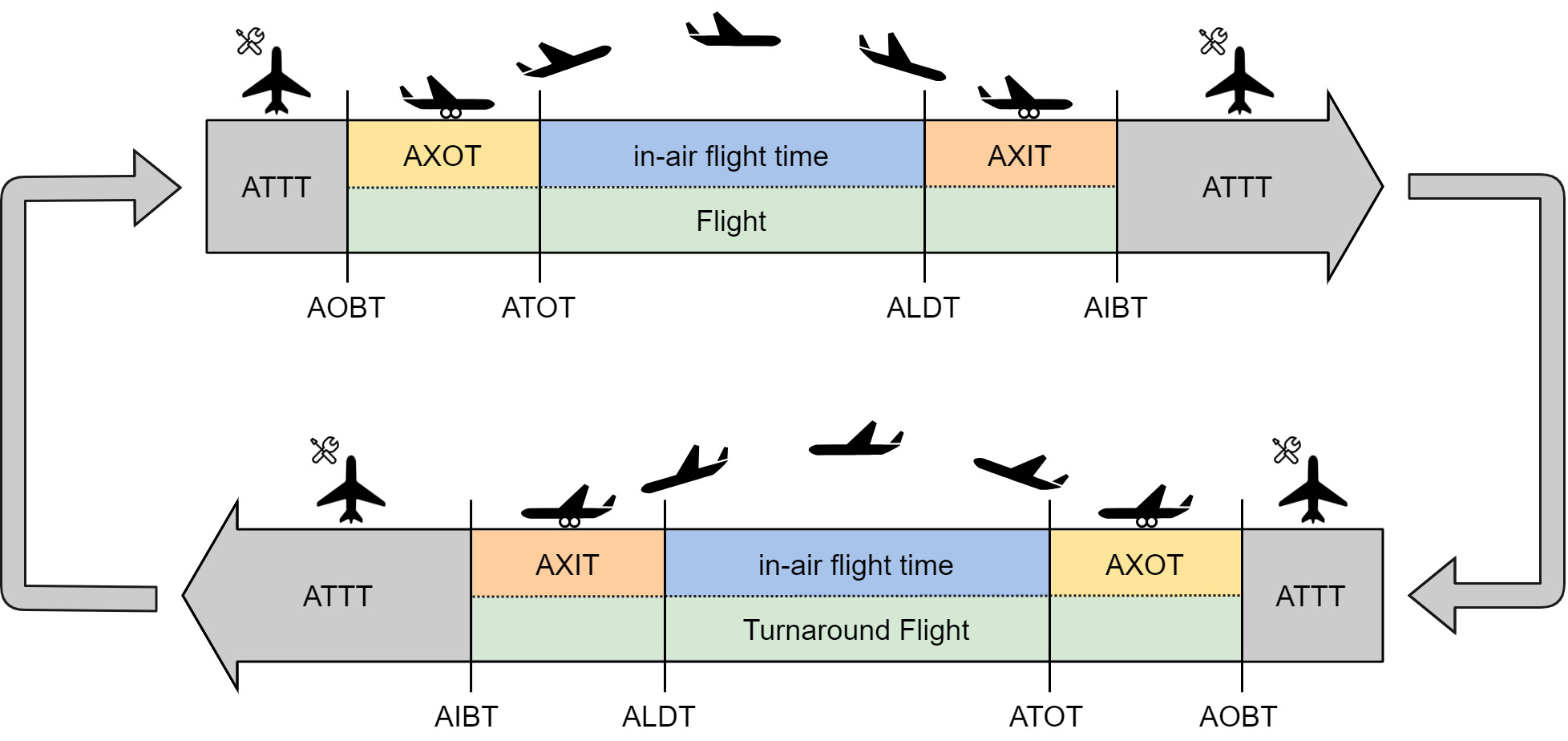}
    \caption{Flight and turnaround flight time information}
    \label{fig:flight_times}
\end{figure*}

The centrepiece of the SmartAeronautics domain is the Flight data model~\footnote{\url{ https://github.com/smart-data-models/dataModel.Aeronautics/tree/master/Flight}}. An entity of type flight represents all the activities related with a regular flight, since it has come out from its parking position in the departure airport until it has arrived to its parking position in the arrival airport. A flight is identified by its numeric flight number. However, it is very extended to concatenate the operator designator (i.e., IATA or ICAO airline code) with the numeric flight number and an optional one-letter suffix. Other attributes that characterise a flight are the state of the flight (i.e., scheduled, active, unknown, redirected, landed, diverted, cancelled); the number of passengers; the destiny and departure airport; the airline designator; and datetime information. Among datetime information, the model differences between general data presented to passengers (e.g., departure and arrival datetime); and datetime information for the operation of the flight following the nomenclature defined in the Airport Collaborative Decision Making (Airport CDM) manual \cite{acdm_eurocontrol}. The Airport CDM is a reference manual elaborated by Eurocontrol that aims to optimise the operations and collaboration among airports. Figure~\ref{fig:flight_times} presents a flight with its common datetime information. It includes the Off-Block Time (datetime when the aircraft moves from its parking position to the runway), Take-Off Time (datetime when the aircraft takes off), Landing Time (datetime when the aircraft lands), and In-Block Time (datetime when the aircraft goes to its parking position from the runway). Depending on the context where they are used, these times are tagged as `Scheduled', `Estimated', `Actual', `Calculated', or `Target'. All of these datetimes are timestamps that delimitate the different phases of any flight. In this sense, the in-air flight time is the time between the Actual Take-Off Time (ATOT) and the Actual Landing Time (ALDT). The complete flight time, known as block-to-block time, takes from the Actual Off-Block Time (AOBT) to the Actual In-Block Time (AIBT). The time slot between the AOBT and the ATOT is known as Actual Taxi-Out Time (AXOT); the time slot between the the ALDT and AIBT is the Actual Taxi-In Time (AXIT); and the time between the AIBT of a flight and the AOBT of the next turnaround flight is the Actual Turn-round Time (ATTT). Figure~\ref{fig:flight_times} shows a summary of all datetimes used in a flight and their official abbreviations. Table~\ref{table:flight_model} collects the properties and relationships of the Flight smart data model with an example.

\begin{table*}
\centering
    \begin{threeparttable}
        \caption{Flight smart data model. Properties and relationships.}
        \begin{tabular}{p{4cm}p{6cm}p{4cm}} \hline
 Field & Description & Example \\ \hline
 id & Flight Identifier & urn:ngsi-ld:Flight:flight-1234 \\
type & NGSI-LD type & Flight \\
flightNumber & Flight number without airline identifier & 1234 \\
flightNumberIATA & Flight number with IATA airline designator & SN1234 \\
flightNumberICAO & Flight number with ICAO airline designator & BEL1234 \\
state & Flight state. One of:scheduled, active, unknown, redirected, diverted, cancelled & active \\
passengerCount & Number of passengers & 12 \\
dateDeparture & Departure datetime of the flight & 2021-02-04T10:40:01.00Z \\
dateArrival & Departure datetime of the arrival & 2021-02-04T12:40:01.00Z \\
date(S/E/A/T)OBT \textsuperscript{a} & Off-Block Time & 2021-02-04T10:40:01.00Z \\
date(E/A/C/T)TOT \textsuperscript{a} & Take-Off Time & 2021-02-04T10:45:01.00Z \\
date(E/A/T)LDT \textsuperscript{a} & Landing Time & 2021-02-04T12:35:01.00Z \\
date(S/E/A)IBT \textsuperscript{a} & In-Block Time & 2021-02-04T12:40:01.00Z \\
date(E/A)XOT \textsuperscript{a} & Taxi-Out Time & 300 \\
date(E/A)XIT \textsuperscript{a} & Taxi-In Time & 300 \\
date(S/E/A)TTT \textsuperscript{a} & Turn-round Time & 1800 \\
hasAircraft & Reference to the flight aircraft & urn:ngsi-ld:Aircraft:aircraft-AAAAA \\
hasAircraftModel & Reference to the flight aircraft model & urn:ngsi-ld:AircraftModel:aircraftModel-AirbusA310-200 \\
departsFromAirport & Reference to the departure airport & urn:ngsi-ld:Airport:airport-BMA \\
arrivesToAirport & Reference to the arrival airport & urn:ngsi-ld:Airport:airport-MAD \\
belongsToAirline & Reference to the flight airline & urn:ngsi-ld:Airline:airline-SN \\ \hline
            \end{tabular}
    \begin{tablenotes}
      \small
      \item{\textsuperscript{a}One property for each character within parentheses, for example: date(E/A)XOT would result in dateEXOT and dateAXOT. S = Scheduled, E = Estimated, A = Actual, C = Calculated, T = Target.}
    \end{tablenotes}
\label{table:flight_model}
    \end{threeparttable}
\end{table*}

Many tasks and events happen in any flight and they have to be registered. The FlightNotification smart data model~\footnote{\url{https://github.com/smart-data-models/dataModel.Aeronautics/tree/master/FlightNotification}} defines this kind of information. A typical flight notification includes a description of the event, the datetime when it was issued, the datetime when it was modified, and the person or machine which made the notification. In the case that the notification represents a task, it also includes its state which could be `active', `inactive', `completed', or `unknown'.

An aircraft entity represents the physical aircraft that operates the flight. It is identified by its registration or tail number. This registration mark includes information about the country where the aircraft is registered and a code associated with the aircraft. The Aircraft entity~\footnote{\url{https://github.com/smart-data-models/dataModel.Aeronautics/tree/master/Aircraft}} models a physical aircraft with real time data including parameters such as its position, speed, heading degrees, or the datetime of their last update. The information of an aircraft is completed with the properties of its aircraft model. The AircraftModel~\footnote{\url{https://github.com/smart-data-models/dataModel.Aeronautics/tree/master/AircraftModel}} includes the physical characteristics that have in common all aircrafts of the same type. Among these characteristics, there is the IATA and ICAO code, length, wingspan, height, maximum speed, etc.   

An Airline entity~\footnote{\url{https://github.com/smart-data-models/dataModel.Aeronautics/tree/master/Airline}} represents a real airline identified by its IATA or ICAO designator code or callsign (i.e., telephony designator). The model includes additional information of the airline such as its name, short name, or country address.

Lastly, the Airport entity~\footnote{\url{https://github.com/smart-data-models/dataModel.Aeronautics/tree/master/Airport}} is a representation of a real airport. An airport is identified by its IATA or ICAO designator and completed with additional information such as its location, address, name, etc.

\begin{figure*}[]
    \centering
    \includegraphics[scale=0.05]{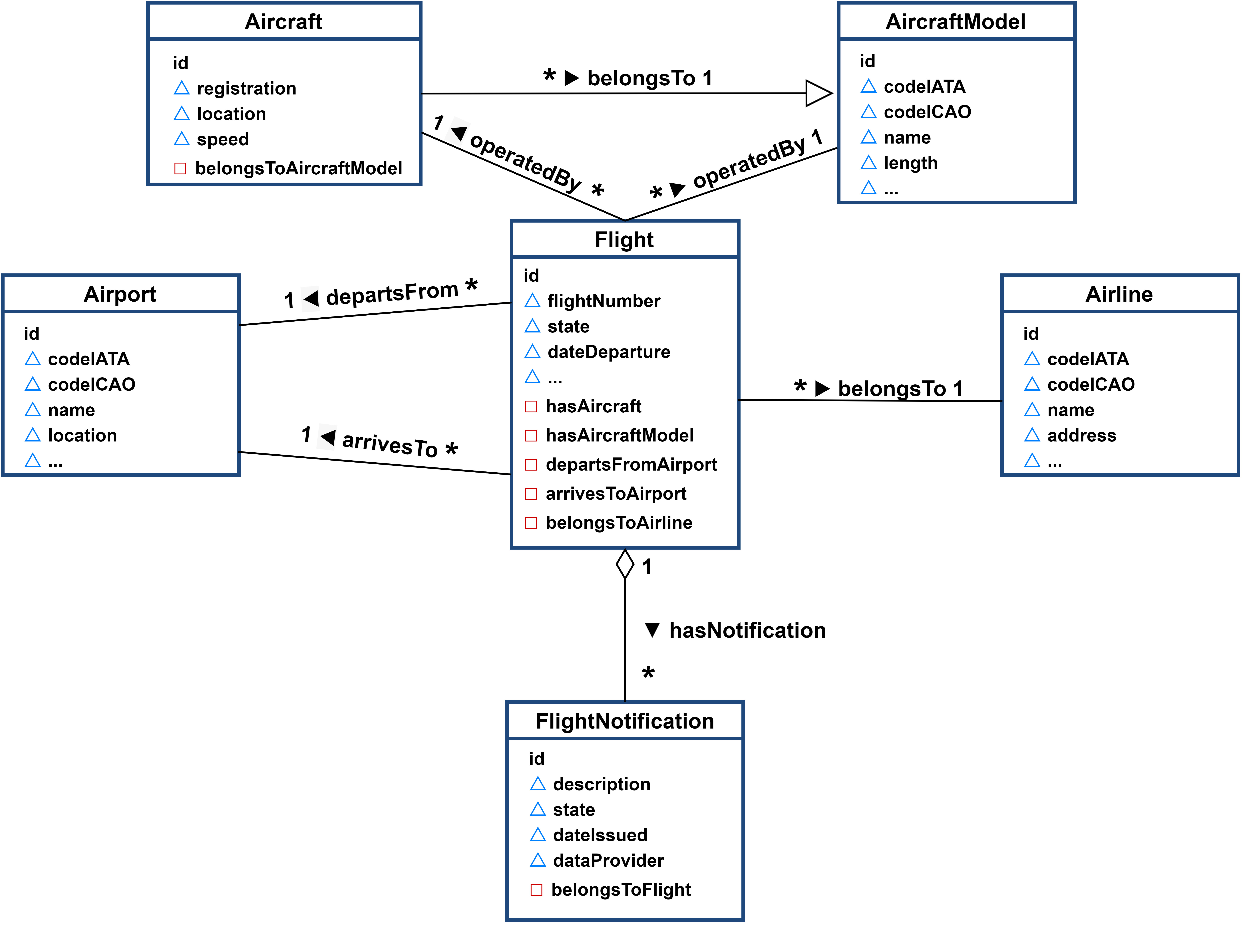}
    \caption{Smart Aeronautics Models}
    \label{fig:aeronautics_model}
\end{figure*}

Figure~\ref{fig:aeronautics_model} summarises the models that involve the aeronautics domain with the relationships among them.

All data models presented have followed the road map defined in the Smart Data Models life-cycle. Firstly, the data models rely under development hosted in the incubated GitHub repository\footnote{\url{https://github.com/smart-data-models/incubated}}(incubation state). Then, the data models have been completed, approved by the FIWARE community, and officially published to the new SmartAeronautics domain. Nowadays, they are available in the FIWARE Smart Data Model catalogue, and their adoption is recommended by the FIWARE community for the development of smart applications and DTs in the domain of aeronautics. However, the data models are not definitively closed, therefore they can be extended by the community through the harmonisation GitHub repository\footnote{\url{https://github.com/smart-data-models/harmonization}} staying in the harmonisation state.

\subsection{Airport DT Data Management}

The previous section defines the data models used in the aeronautics case for representing the different entities that will be available for consumers. Thanks to the defined data models, those consumers (typically applications and services) are able to obtain standardised data with an homogeneous structure. However, as stated before, one of the main challenges of data management and digital twin scenarios is related to the process of obtaining data from different data sources and adapting them to be compliant with the defined models.

\subsubsection{Obtaining data from external sources}

Obtaining and merging data from external sources represents a challenge because usually data come from sources of a very different nature. We can find sources that expose the data in different formats, with different structures, using different naming for referring to the same concepts, providing them with different frequencies, etc. 

Therefore, the first step in the process of data management consists on the homogenisation of the available data for fitting the defined data models. \added{In the parking use case, data came from IoT sensors and REST APIs homogenised by the FIWARE IoT Agents and the Draco GE respectively. In the airport DT architecture, we use Draco to perform these tasks. The main difference among DTs implemented using the FIWARE technology is the configuration of Draco processors to obtain and homogenise the information, which is totally dependent on the data sources.} Specifically, in the use case presented in this manuscript, we use the following two data sources for obtaining airport data: Azinq Chroma API~\footnote{\url{https://azinq.com/}} and Plane Finder~\footnote{\url{https://planefinder.net/}}.  

Chroma feeds the Digital Twin with information about flights, airports, and airlines. The service provides a REST API that is established through an HTTPs polling system, providing the client its credentials in every request. A Draco processor is in charge of establishing the connection and periodically sending HTTP requests to the service. 

The received information flows in Draco through a set of configurable and reusable processors which develop simple operations over the data to achieve the format specified in the data model~\footnote{\url{https://nifi.apache.org/docs.html}}\textsuperscript{,}~\footnote{\url{https://fiware-draco.readthedocs.io/processors\_catalogue/introduction/index.html}}. The specific processors configured in Draco are explained below. We have included an example to illustrate how data is processed with them: 

\begin{itemize}
    \item InvokeHTTP-Input-Flight : it makes the HTTPS requests to the flight endpoint of Chroma API with a run schedule (polling frequency) of one minute. Listing~\ref{lst:chroma} shows an example of data received from the API.
    
\begin{lstlisting}[caption={Example of data received from Chroma},label={lst:chroma},captionpos=b,language=json,basicstyle=~\footnotesize]
[
  {
    "id": 0,
    "ScheduledDateTime": null,
    ....
  },
  {
    "id": 1,
    "FlightNumber": "1234",
    "AirlineIATA": "SK",
    "DepartureArrivalType": "A",
    "OriginDestAirportIATA": "SVG",
    "OriginDestAirportICAO": "ENZV",
    "Registration": "AAAAA",
    "StandCode": "01",
    "GateCode": "01",
    "ALDT": null,
    "AIBT":  "2021-02-04T17:20:00+00:00",
    "AOBT": null,
    "TOBT": null,
    "ScheduledDateTime": "2021-02-04T17:20:00+00:00"
  }
]
\end{lstlisting}

    \item SplitJson: it segregates all incoming flight entries into individual ones. In our example, it divides the array into two new entries: the flight with `id' 0 and the flight with `id' 1.
    \item EvaluateJsonPath: it takes some entity attributes from the entries for future processing in the Draco processor chain. Concretely, in the case of flight entities, it takes the `DepartureArrivalType' and `OriginDestAirportIATA' fields to build the arrival and departure airport properties. 
    \item RouteOnAttribute: it filters all flights with `ScheduledDateTime' not null. In our example, it deletes the flight with `id' 0.
    \item UpdateAttribute: it builds new attributes. It creates the `arrivesToAirport' and `departsFromAirport' attributes from the original `DepartureArrivalType' and `OriginDestAirportIATA' fields. The logic of the processor does the following: if the `DepartureArrivalType' is marked as `A' (arrival), then the `arrivesToAirport' will be `ABZ' (Aberdeen’s airport) and `departsFromAirport' will be the `OriginDestAirportIATA'. In contrast, if the `DepartureArrivalType' is marked as `D' (departure), then the `arrivesToAirport' will be `OriginDestAirportIATA' and the `departsFromAirport' will be `ABZ' (Aberdeen’s airport).
    \item JoltTransformJson\_To\_NGSI: it transforms the input flight entity to be compliant with the NGSI-LD format. The
    processor is based on the Jolt Java library included in NiFi that allows to transform an input JSON into an output JSON through a specification. The result of the transformation of flight 1234 is shown in Listing~\ref{lst:chroma_transformed}.
    \item ReplaceText\_Sanetize: it sanitises the JSON to avoid characters not allowed in NGSI-LD format.

    \item InvokeHTTP\_POST\_Context\_Broker: it makes an HTTP POST request saving the flight entity in the Context Broker.
\end{itemize}

\begin{lstlisting}[caption={Chroma API data transformed to NGSI-LD Flight Entity},label={lst:chroma_transformed},captionpos=b,language=json,basicstyle=~\footnotesize]
{
  "id": "urn:ngsi-ld:Flight:flight-1",
  "type": "Flight",
  "flightNumber": {
    "value": "1234",
    "type": "Property"
  },
  "belongsToAirline": {
    "value": "urn:ngsi-ld:Airline:airline-SK",
    "type": "Relationship"
  },
  "departsFromAirport": {
    "value": "urn:ngsi-ld:Airline:airport-SVG",
    "type": "Relationship"
  },
  "arrivesToAirport": {
    "value": "urn:ngsi-ld:Airline:airport-ABZ",
    "type": "Relationship"
  },
  "hasAircraft": {
    "value": "urn:ngsi-ld:Aircraft:aircraft-AAAAA",
    "type": "Relationship"
  },
  "standCode": {
    "value": "01",
    "type": "Property"
  },
  "gateCode": {
    "value": "01",
    "type": "Property"
  },
  "dateAIBT": {
    "value": {
      "@type": "DateTime",
      "@value": "2021-02-04T17:20:00.00Z"
    },
    "type": "Property"
  },
  "dateScheduled": {
    "value": {
      "@type": "DateTime",
      "@value": "2021-02-04T17:20:00.00Z"
    },
    "type": "Property"
  },
  "@context": [
    "https://smartdatamodels.org/context.jsonld",
    "https://uri.etsi.org/ngsi-ld/v1/ngsi-ld-core-context.jsonld"
  ]
}
\end{lstlisting}

\begin{figure*}[]
    \centering
    \includegraphics[scale=0.07]{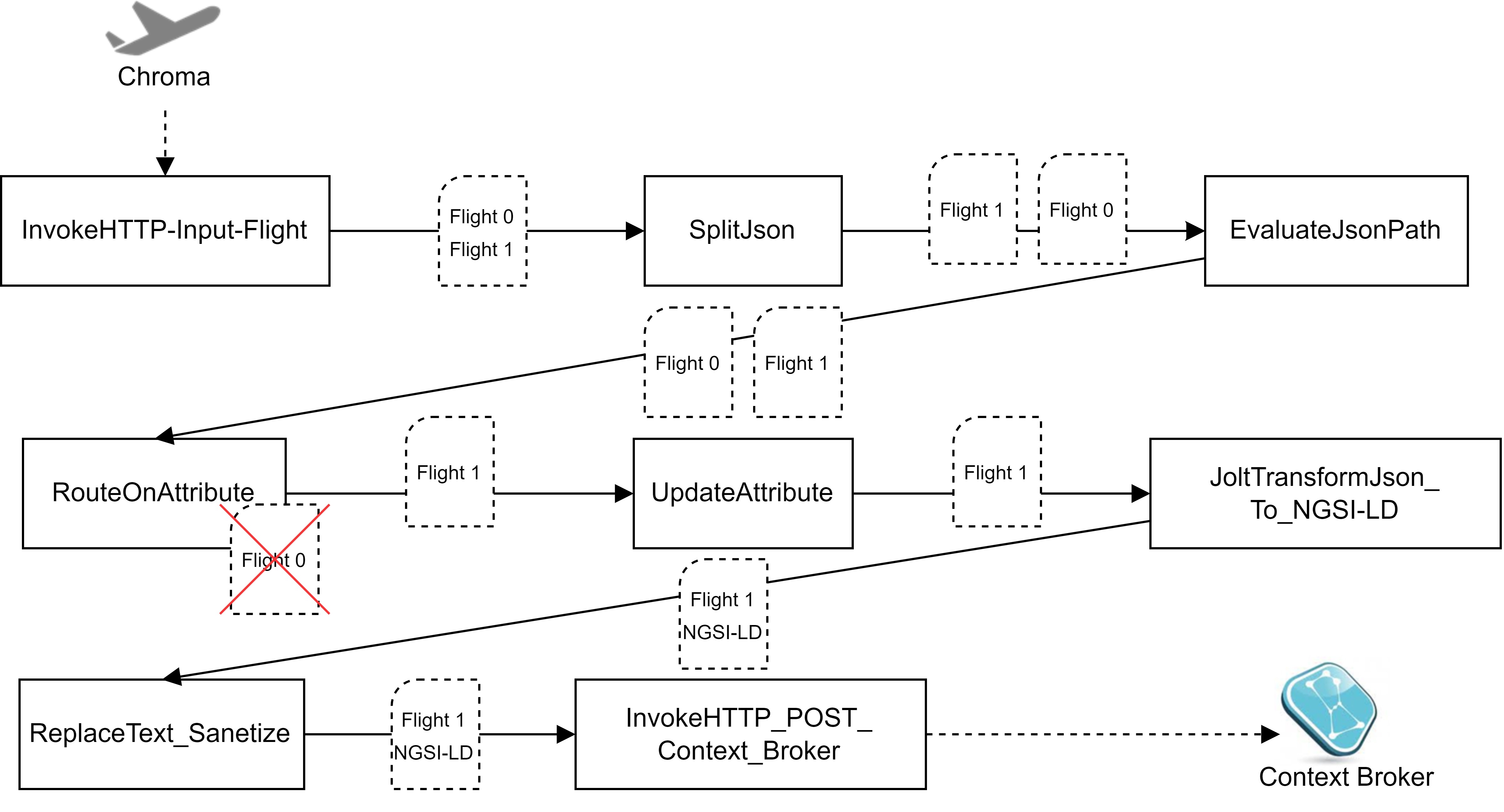}
    \caption{\added{Draco processors connection for capturing data from Chroma API}}
    \label{fig:draco_chroma}
\end{figure*}

Figure~\ref{fig:draco_chroma} summarises the chain of processors involved in data acquisition from Chroma API. The templates for airports and airlines are very similar to the flight one, configuring each processor adapted to those endpoints.

The other data source is Plane Finder. It is an API that feeds the Digital Twin with information about the position of aircrafts located near the airport and provided by stations located around the airport. The connection to this service is established through a TCP socket secured with TLS1.2 (\added{Transport Layer Security}). When the handshake is completed, the client receives data updates through a dedicated channel for a long period of time.

In this case, the Draco processors that transform the data into the defined format are the following: 

\begin{itemize}
    \item TCPClient: it establishes the connection with Plane Finder Firehouse.
    \item DecompressContent: it decompresses the received data. Listing~\ref{lst:plane_finder} shows an example of decompressed data provided by Plane Finder.

\begin{lstlisting}[caption={Example of data received from Plane Finder},label={lst:plane_finder},captionpos=b,language=json,basicstyle=~\footnotesize]
{
  "X": {
    "reg": "AA-AAAA",
    "flight_number": "SK1234",
    "adshex": "X",
    "lat": 57.305525,
    "lon": -1.622521,
    "altitude": 7675,
    "heading": 222,
    "speed": 281,
    "vert_rate": -1856,
    "is_on_ground": false,
    "pos_update_time": 1612457454,
    "route": "SVG-ABZ"
  },
  "Y": {
    "route": "SVG-AXZ",
    ...
  }
}

\end{lstlisting}

    \item SplitJson:  it segregates all incoming aircrafts into individual ones. In the case above, it divides the JSON into two new JSONs, one containing the aircraft X and the other containing the aircraft Y.
    \item EvaluateJsonPath: it gets some entity attributes for future processing. Concretely, it takes the `flightNumber', `pos\_update\_time' (timestamp of the measure), `reg' (register number of the aircraft), and `route'. 
    \item RouteOnAttribute: it selects all aircrafts whose flights depart from or arrive at Aberdeen’s Airport and whose `flightNumber' is not null.
    \item UpdateAttribute: it builds some attributes related to the aircraft entity. It adapts the `flightNumber' to be the same as provided by the Chroma API, i.e., without including the IATA code of the airline. It also modifies the `reg' field deleting the ‘`-' character because the register provided by Chroma does not include it. Lastly, it creates a new attribute called `dateIssued' from the `pos\_update\_time', transforming it to \added{International Organization for Standardization} (ISO) 8601 datetime format, the one recommended by the data models.
    \item JoltTransformJson\_To\_NGSI: it builds an aircraft entity with NGSI-LD format. Listing~\ref{lst:plane_finder_transformed} shows the resultant aircraft entity. This new entity represents the position of the aircraft using the \added{Geographic JSON} (GeoJSON) format and it updates the incoming data with the unit of measure proposed in the data model (e.g., it transforms `altitude' from feet to metres).
    \item ReplaceText\_Sanetize: it sanitises the JSON deleting special characters not admitted by NGSI-LD.
    \item InvokeHTTP\_POST\_Context\_Broker: it makes an HTTP POST request saving the aircraft entity in the Context Broker.
\end{itemize}

\begin{lstlisting}[caption={Plane Finder data transformed to NGSI-LD Aircraft Entity
},label={lst:plane_finder_transformed},captionpos=b,language=json,basicstyle=~\footnotesize]
{
  "id": "urn:ngsi-ld:Aircraft:aircraft-AAAAA",
  "type": "Aircraft",
  "flightNumber": {
    "value": "1234",
    "type": "Property"
  },
  "flightNumberIATA": {
    "value": "SK4615",
    "type": "Property"
  },
  "adshex": {
    "value": "X",
    "type": "Property"
  },
  "location": {
    "value": {
      "type": "Point",
      "coordinates": [
        57.305525,
        -1.622521,
        2339.339925
      ]
    },
    "type": "GeoProperty"
  },
  "heading": {
    "value": 222,
    "type": "Property"
  },
  "speed": {
    "value": 520.411811,
    "type": "Property"
  },
  "verticalSpeed": {
    "value": -9.428499,
    "type": "Property"
  },
  "isOnGround": {
    "value": false,
    "type": "Property"
  },
  "dateIssued": {
    "value": {
      "@type": "DateTime",
      "@value": "2021-02-04T16:50:54.00Z"
    },
    "type": "Property"
  },
  "@context": [
    "https://smartdatamodels.org/context.jsonld",
    "https://uri.etsi.org/ngsi-ld/v1/ngsi-ld-core-context.jsonld"
  ]
}

\end{lstlisting}

With these processors, we have validated that FIWARE Draco addresses the problem of data fusion and homogenisation, two  of the most common challenges when developing DTs Fuller et al. \cite{9103025}. We tested that with FIWARE Draco it is possible to merge data from (1) different sources, (2) with different formats, (3) using different protocols, and with (4) different update periods. Moreover, it allows to work in Big Data environments. Instead of coding a program that makes all the transformations the problem is divided and solved with processors that implement simple functions. The advantage of this scope is that these processors can be reusable. 

\subsubsection{Providing data to consumers}

\begin{figure*}
    \centering
    \includegraphics[scale=0.15]{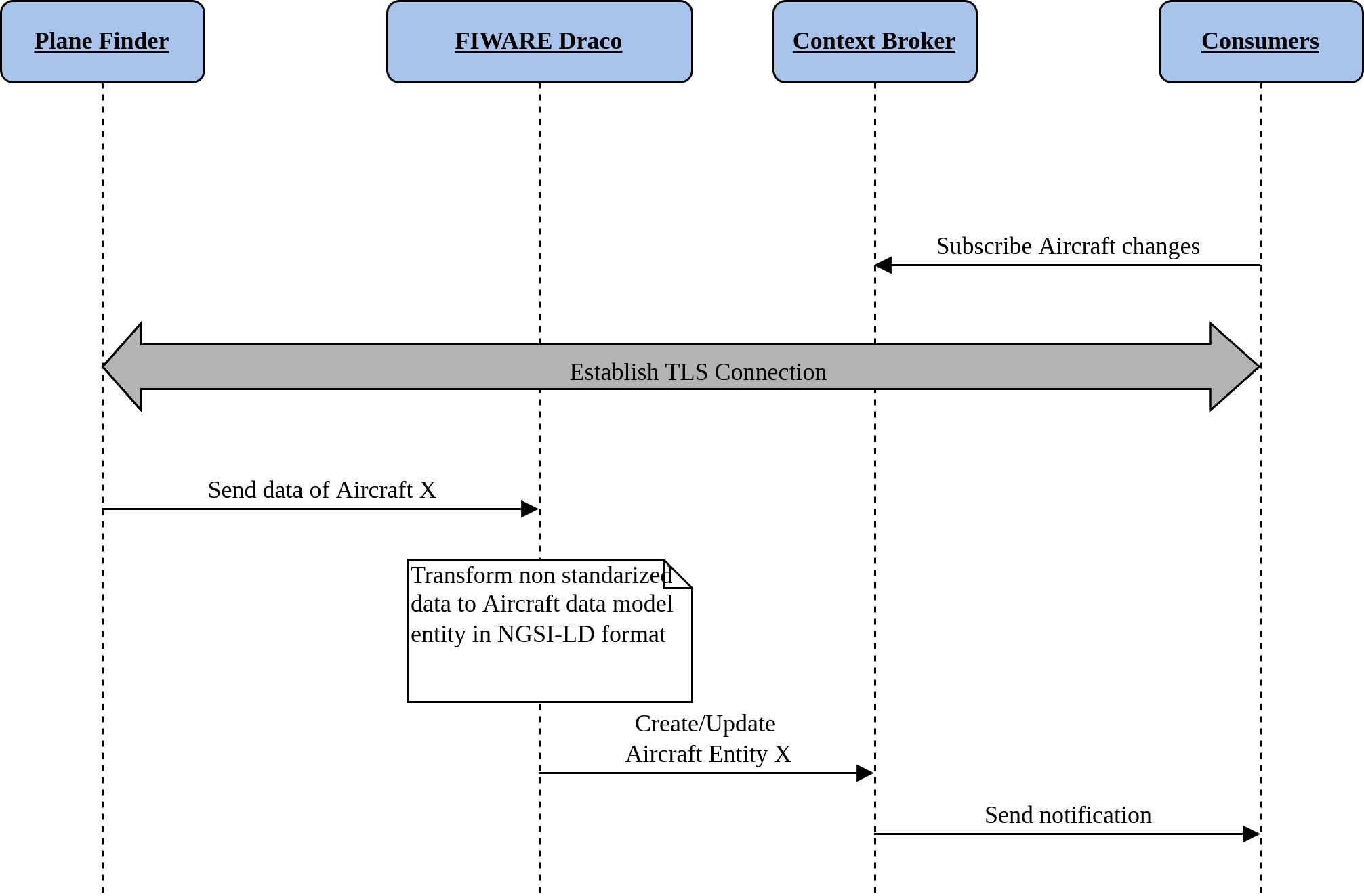}
    \caption{Example of asynchronous communication}
    \label{fig:async_com}
\end{figure*}

The communication with consumers is established through the Context Broker, which is the one that manages all the operations. Consumers not only read data, they can also update or create new entities. The process of accessing data follows two approaches. On the one hand, there is an asynchronous communication, where clients, who previously sent a subscription to the Context Broker, receive HTTP notifications when the entity (or a property of an entity) they were subscribed has changed. This kind of communication allows the consumers to remain updated without the requirement of making periodical requests to the Context Broker. In the Aberdeen’s airport DT, these subscriptions are configured to receive changes in flights. Figure~\ref{fig:async_com}. shows how asynchronous communication works with an example of an update in an aircraft entity.  

On the other hand, there is a synchronous communication where clients directly send HTTP requests to the Context Broker for creating, reading, updating, or deleting entities, for example, getting information of flights, or creating, flight notifications. Among the most common operations made by consumers are getting a flight by its flight number, filtering flights by date, getting the position of an aircraft, and creating, reading, updating, or deleting flight notifications.

\added{The parking DT use case presented the same communication mechanisms to serve information to consumers. The main difference is in the implementation of consumer applications. While the parking DT implemented a mobile application to manage a parking, the airport DT presents a system to manage flight turnaround operations}.

Lastly, the solution presented also allows to get the historical data. However, these data are not managed by the Context Broker, because the Context Broker only saves the current state, i.e., the last updated value of the entities. Instead, the historical data are managed by Draco which saves them in a Mongo database, making them accessible for consumers.


\section{Data visualisation}
\label{sec:visualization}

The definition of DTs proposed by Tao et al. \cite{TAO2018169} (extending the dimensions collected by Grieves \cite{grieves2014digital}), includes external services that consume the processed data as an important actor in the developments of DTs. This section explains how the data managed by our reference architecture, presented in the previous section, are consumed by the Airport Digital Twin. The objective of this use case is to reduce the delays in flights. For that purpose, two client applications communicate with the rest of the architecture of the DT to ease the work of operation managers. Both applications are based on web technologies. The first one is a desktop application that shows the real state of the airport combining 2D and 3D virtual reality (\added{VR}) representations of the stands. The second one is a mobile application addressed to ground operators. It consists in a dispatcher system that allows the operators to register and read events about the aircraft during the turnaround stage.

\subsection{Airport 2D/3D digital representation}

The 2D/3D digital representation of the airport will be used by the control room as a mechanism to analyse the real state of the airport and to assign tasks to airport operators in an efficient way. It is an interactive application that allows users to navigate along the airport model which updates itself through the notifications received by the Context Broker. 

\begin{figure}
    \centering
    \includegraphics[scale=0.065]{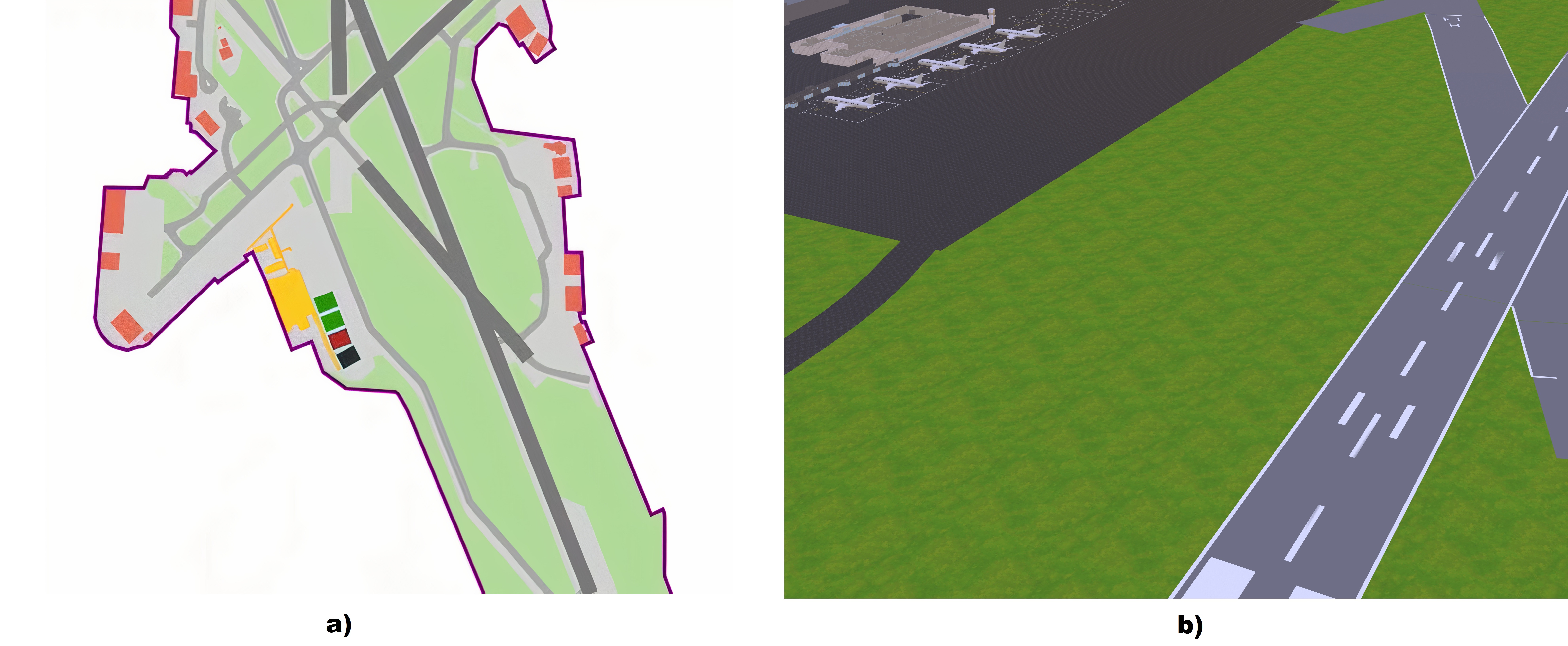}
    \caption{\added{Virtual representations of Aberdeen International Airport. (\textbf{a}) 2D virtual representation. (\textbf{b}) 3D virtual representation.}}
    \label{fig:virtual_representation}
\end{figure}

During the design phase of the application, three main concerns were taken into consideration: (1) it should be as platform-agnostic as possible, as technological equipment is likely to change over the years; (2) it must be easy to integrate new functionalities for both the 2D and 3D representations systems; (3) VR capabilities have to be granted to help workers to navigate through the system. With the aim of solving the first issue, a web application was decided as the client for the tool. In this way, the same code could be used along different computers, browsers, and devices. It also means that the application could be accessed from any machine by just using the operator credentials and delegating the security and logic to the server side. There are multiple architectures to implement view and controller patterns. However, due to a strict need of security, each view was served independently so each time the user’s credentials could be checked again. This led to a static-page system in which content is requested just after the web is loaded in the user. Regarding the second and third requirements, it is necessary to use a library that enables the integration of virtual reality in web browsers; it must be easy to extend the solution; it must be efficient as it needs to work with big complex models; and it has to be actively maintained, being compliant with the new versions of JavaScript and web browsers. In this sense, two libraries were taken as main candidates: Three.js~\footnote{\url{https://threejs.org/}} and Babylon.js~\footnote{\url{https://www.babylonjs.com/}}. Among the differences between them, one of the most remarkable ones is that Three.js is based on \added{HyperText Markup Language} (HTML) tags to build a scene, while Babylon.js mainly works with pure JavaScript and a reference to a canvas element to draw a scene. Due to this dynamic approach, more powerful VR capabilities, and a large set of examples and documentation, Babylon.js was finally chosen. Moreover, the validation of Babylon.js for DTs has been probed in other researches such as the one conducted by Falah et al. \cite{FALAH2020331}. Figure~\ref{fig:virtual_representation} shows the resulting 2D and 3D representations of the Aberdeen International Airport accessible through the web application.

\subsection{Dispatcher system}

The dispatcher system is a mobile application used by ground operators to manage turnaround events. Its objective is to improve decision making, following the Airport Collaborative Decision Making standard \cite{acdm_eurocontrol} and helping airports integrate it \cite{OKWIR2017183}. Typically, operations managers deal with vast amount of information which is dispersed in different and heterogeneous formats and systems. The dispatcher places all the necessary information ready at hand, at all the places, and for all the people involved in operations, so that decisions can always be based on complete information. This requires significant effort in collecting, storing, and distributing data in real time, so a suitable architecture is required to meet these requirements, as the one presented in Section~\ref{sec:architecture}.

We developed the application through a number of user-centred design cycles and participatory design sessions where we performed focus groups, interviews, design sessions, and formative evaluations with increasing fidelity prototypes to afford grounded discussions. In total, we held five meetings where participated seven stakeholders from the staff at the collaborating airport. We worked with operations personnel, including a ramp manager, air side operations supervisors and managers, airport duty managers and handling agents. Table~\ref{table:participants} provides details and descriptions of the stakeholders in the participatory design sessions. From these meetings some conclusions were drawn related to the objectives of the system: 

\begin{itemize}
    \item \textbf{Efficient use of stands and gates}: Optimisation of the use of the stand and the gate. Operators are restricted in the number of gates they can use simultaneously. It is important to use the right gate for the right aircraft in the right periods. Participants aim to optimise the usage of gates and the turnaround process. Efficient use of stands and gates will produce positive results for a number of factors, including fuel savings and departure times.
    
    \item \textbf{Complete information for decision making}: Operations managers look at efficiency in decision making. Often, they need to change the programme of events after an unplanned event. They need to adapt. Typically, operations managers do not have all the information in one place. Often, they lack information that is relevant to the entire operation. A goal of the system is to place all the necessary information ready at hand in all places and for all the people involved in the operations, so that decisions can always be based on complete information. This requires a significant effort to collect, store, and distribute data in real time.
    
    \item \textbf{Efficient passenger journey}: Operations managers express the need to reduce as much as possible the distances passengers need to travel from gate to gate. They aim to reduce travel time and the possibilities of getting lost. This is particularly critical during last-minute changes in stand and gate numbers.
\end{itemize}

The first meeting aimed at capturing current roles, tasks, and data sources. The group also decided that a mobile application supporting the dispatcher in entering and reading information about the aircraft during its turnaround should be the design goal. Between the first and second session we designed the first low-fidelity prototype described in the section below. During the second session, we ran a formative evaluation and a live redesign exercise to improve the tools in supporting concrete tasks, primarily taking into consideration the standards and conventions held by the personnel. The third and fourth sessions reviewed the second and third prototypes respectively. The third prototype was the last low-fidelity prototype before a fully-functioning prototype was developed and tested during the fifth and final session.

\begin{table*}
\centering
\caption{Participants’ descriptions and data tools.}
\begin{tabular}{p{1cm}p{7cm}p{9cm}}
\hline
 \textbf{Id} & \textbf{Position} & \textbf{Tasks} \\ 
 \hline
P1 & Ramp manager & Managing the ramp team, providing support and overview of all departments and tasks \\
P2 & Airside Operations Supervisor & Maintaining safety and compliance in all aspects of the airfield  \\
P3 & Airside Operations Manager &  Ensuring continuity/integrity of air side operations  \\
P4 & Airport Duty Manager & Operational management of resources, both asset and people \\
P5 & Airport Duty Manager (Chroma sysadmin) & Planning operational scenarios on fixed resources i.e. Check-in, Baggage Carousel, Gates and Stands \\
P6 & Head of Innovation and Continuous Improvement & Project Management for Airport Twin 2020 \\
P7 & Handling agent & Taking care of everything that passengers and crew require, including ground transport, executive services, meet and greet, assistance for passengers, etc.  \\ 
\hline
\end{tabular}
\label{table:participants}
\end{table*}

\subsubsection{Operations Flowcharts}

\begin{figure*}
    \centering
    \includegraphics[scale=0.4]{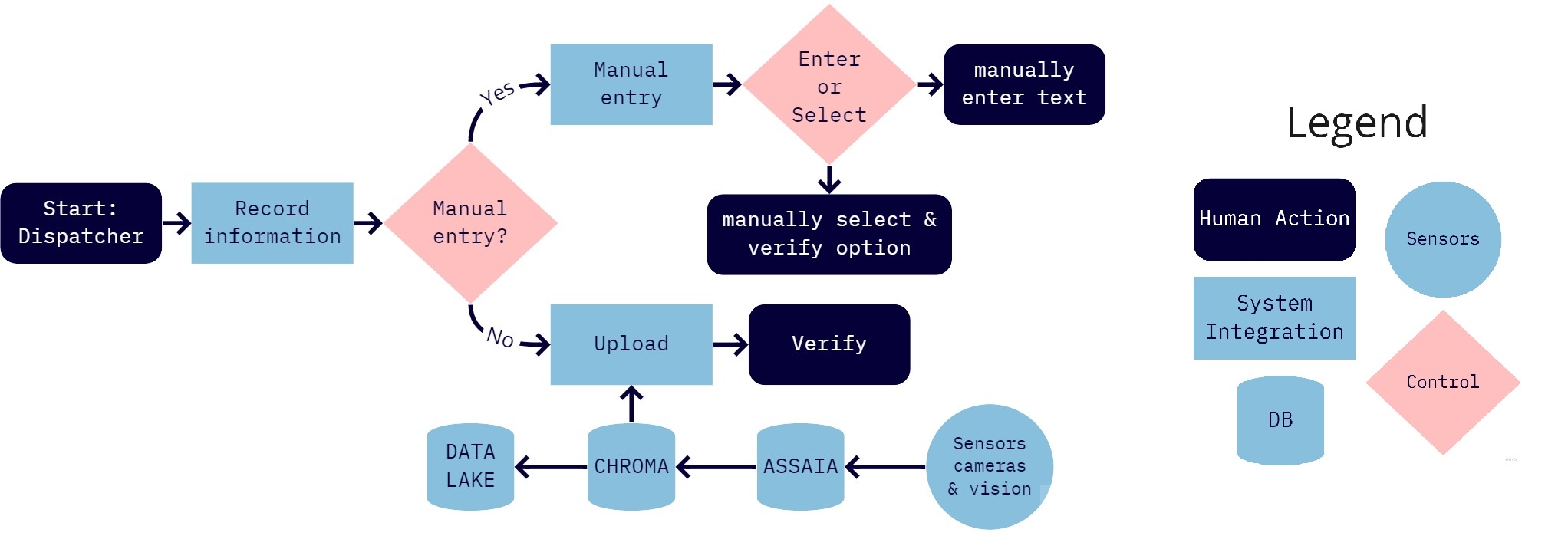}
    \caption{Dispatcher Flowchart}
    \label{fig:dispatcher_flowchart}
\end{figure*}

The discussion with the participants began with drawing the dispatcher flowchart in Figure~\ref{fig:dispatcher_flowchart}. The chart describes the flow of information in the decision making process of dispatching. The dispatcher records information manually or automatically through the Context Broker and Chroma. The information includes computer vision detection of key moments in the aircraft’s turnaround. After Chroma uploads the information to the dispatcher, the Context Broker processes the data and notifies the operators. Afterwards, the operator verifies the information with live observations at the stand. When the information is not uploaded automatically through Chroma, the dispatcher enters the information manually directly to the Context Broker. Before the dispatcher application, some information was gathered through text messages, phone calls, or recorded in paper and pencil, continuing with the lack of digialisation detected by Makhloof et al. \cite{makhloof_real}. There was no central information system resulting in an inefficient time-consuming effort. The mobile input/output device dispatcher application allows operators to record information and to receive notifications, bypassing the paper and pencil, reducing the time required for turnaround operations, and as a consequence, reducing flight delays. 

\subsubsection{Design Cycles}

The design of the dispatcher system was completed through three cycles. The application is composed by four main views, a global view with information of all the stands, two views with detailed information about a specific stand, and a view with information about a specific task. 

\begin{figure*}
    \centering
    \includegraphics[scale=0.05]{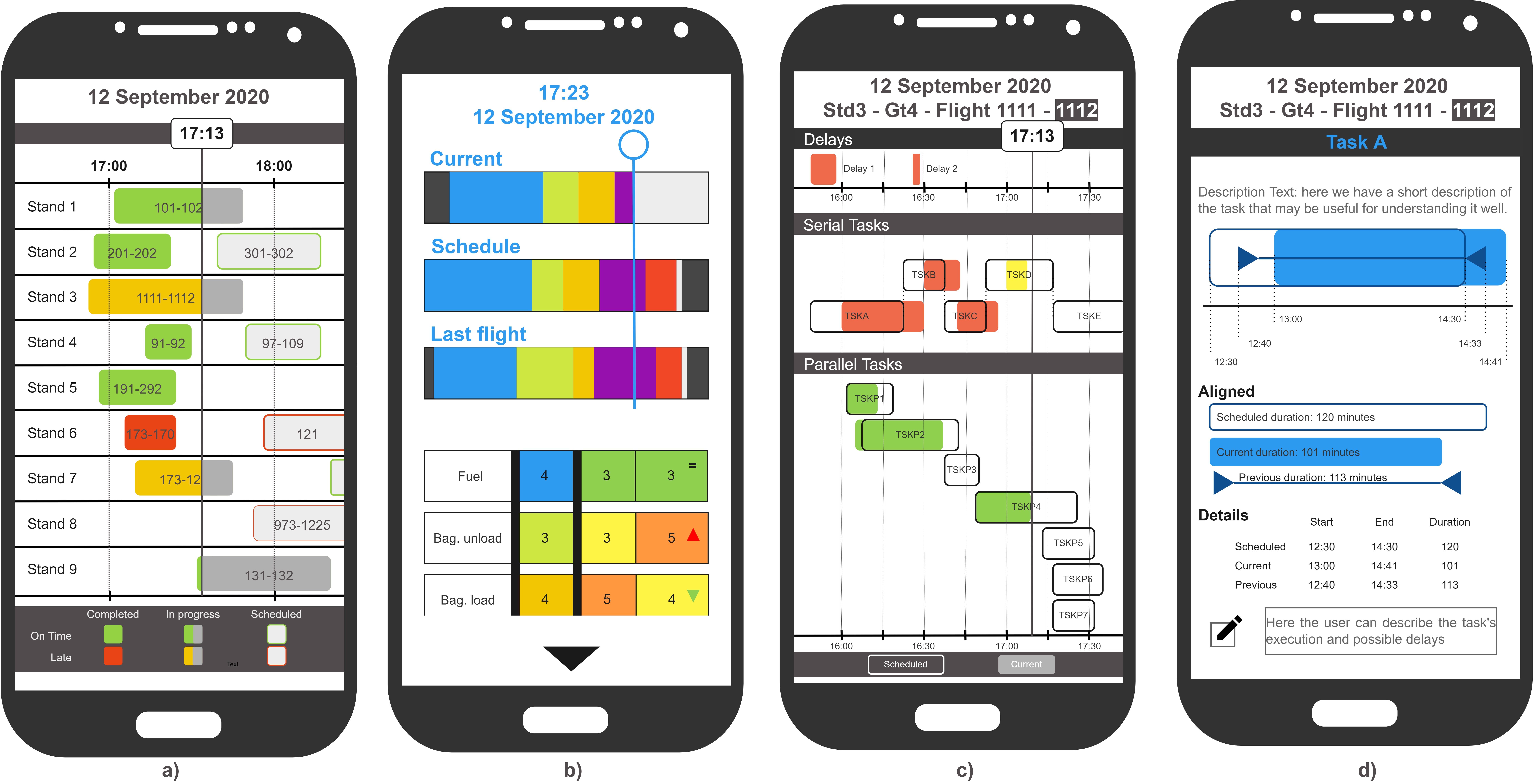}
    \caption{\added{Final version of the DT dispatcher application. (\textbf{a}) Global view. (\textbf{b}) Stand Table View. (\textbf{c}) Stand Gantt view. (\textbf{d}) Task view.}}
    \label{fig:dispatcher_version3}
\end{figure*}

The first design focused on the information and view transfer between the global view and the two versions of the stand view. In the global view, each flight is represented as a colour-coded rectangle in the row encoding its stand. A vertical blue line represents the current time and separates the past (left), from the future (right). Flights that have already begun a turnaround are colour coded based on their adherence to the schedule. Red flights are late, green flights are on time, and so on. Flights in the future are grey. They were developed two stand views, one with the information in a table and the other represented as a Gantt Diagram. The display conveys detailed information on the flight currently in the stand with all the tasks that require completion. Again, a vertical line represents the current time. After discussing and evaluating the limitations of the first version, the participants decided on the improvements for the second one. In this version, the global view was simplified, including a simpler colour coding to represent flight delays. Moreover, a new view with detailed information about a specific task was added. In the final prototype, we included a flight information display in the gate view and we aligned the colour map with the standards used at the airport. We also simplified the Gantt chart view on the right to clearly display the dependent and parallel subtasks of flight turnarounds. Figure~\ref{fig:dispatcher_version3} shows the final version of the dispatcher mobile application.

\section{ \added{Practical Implications} and extending the solution to other airports}
\label{sec:extension}

\added{ In this work, we have proposed an architecture for developing a DT to manage turnaround operations in airports. The proposal was validated in the Aberdeen International Airport, but it can be extended to other airports. The solution is based on Open Source initiatives and the data are modelled following the standards. It is interesting to analyse the extension of our proposal to other airports. There are more than 40,000 airports all over the world, of which 17,600 are commercial. Thanks to the scalable capabilities of the FIWARE GEs, the presented solution can be applied in small, medium and large airports. From the commercial point of view, the solution presents a unified and generic web-based platform interoperable with extended reality technologies. It also supports the management of real-time and batch data processing and the possibility of being integrated with external data providers.} 

\added{When applying our solution to other airports}, important topics must be addressed, such as the capacity of scaling the solution to larger airports, the required effort to migrate to other airports, the technological requirements for airports with low level of digitalisation, etc.

The data model is fully reusable, as it follows the standards of the industry. Moreover, if required, it can be extended for each airport adding new entities, properties, or relationships. The solution is also scalable and feasible to large airports, as it is based on FIWARE GEs integrated as independent services. Regarding the level of digitalisation required to implement the solution, it is necessary that there exist some sources of data which feed the DT. It is important to ensure the data quality. In our proposal, the data sources were Azinq Chroma API and Plane Finder. However, these sources are not compulsory, they can be replaced by others. This is possible thanks to FIWARE Draco, based on NiFi, in where the data fusion operations are performed. This piece is essential for any DT, as it allows to integrate heterogeneous data sources and to adapt them to the data model. One of the main barriers to expanding the solution is that each airport must develop a new 3D model, although some components like the aircrafts can be reused. \added{To solve this issue some technologies as laser scanning for building 3D models, or photogrammetry (i.e., generate 3D models from 2D images) can be used. The resultant models should be integrated with Babylon.js \cite{8560496}. This technology can be combined with drone systems equipped with specific sensors, such as cameras or Light Detection and Ranging (LiDAR) devices, suitable for the construction of 3D models in large spaces. Moreover, AI techniques can be applied to improve the results obtained in the construction of 3D models \cite{math9233033}.} 


\section{Conclusions and future work}
\label{sec:conclusions}

In this paper, we have presented the experience of applying the Digital Twin paradigm to a real aeronautics case deployed in Aberdeen's airport. As explained before, DTs present an opportunity for managing processes and consuming contextual information in a wide range of scenarios and applications. We have presented a detailed architecture of a DT for managing airport information based on open standards and FIWARE-based technologies. These include a set of open source software components and modelling specifications that used all together allows the development of reusable and comprehensive DT designs. Specifically, the presented architecture takes advantage of Draco, Orion Context Broker, NGSI-LD, and FIWARE Data Models to provide real-time and historical data about airport flights. These data include information about plane positioning, flight scheduling, and turnaround operations. Data life-cycle is entirely managed by the presented architecture, from the data acquisition and their modelling to the storing and provisioning to be consumed by final applications.

Besides the data management architecture, the manuscript describes two real applications that are being used in the airport for exploiting the information provided by the DT, i.e, a 2D/3D digital representation of the airport and a mobile-based dispatcher system used by ground operators to manage the turnaround events. Both applications, together with the DT architecture that makes them usable, have been deployed and tested in Aberdeen's airport and are consuming real flight information by the date of writing this work. 

Finally, we present a discussion about how the specific airport use case can be migrated and applied to other existing airports and how the generic DT architecture can also be adapted to multiple scenarios. We conclude that thanks to the detailed description of the DT design and the use case scenarios we have presented, readers could easily apply the learning concepts for developing DTs in the scope of any existing airport for improving the productiveness and usability of flight management. Moreover, following the proposed design and the building guide published by Conde et al. \cite{9346030}, similar DTs infrastructures can be flexibly designed and deployed in any scenario or domain. 

\added{The main challenge of the application of the proposal to real scenarios is related with the capability of adopting the solution by the operators of the airports. In these kinds of scenarios, in which security and reliability are crucial, modifying traditional operating mechanisms is an important barrier to ensure the success of the proposed solution. On the other hand, the main limitation of the proposal is that it is totally dependent on correct and comprehensive sensorization of the airports as well as on the availability of services and APIs to obtain the information of aircraft locations and turnaround operations.}

For continuing the research in this field, we propose validating the architecture presented with new use cases, for example a system that improves the baggage tracking. We also propose expanding the aeronautics domain, adding more properties and relationships to the existing models, or including new ones, such as the runway, the stand, or the aeroplane walkway. Additionally, we propose validating the solution in other airports, analysing the barriers presented in Section~\ref{sec:extension} and including new ones. Of special interest is to study the scalability of the solution to many airports, including those with little technological development.

\section*{Declaration of competing interest}

The authors report there are no competing interests to declare.

\section*{Acknowledgements}

Authors would line to acknowledge to all the colleagues that have participated in the conception, design, development, and materialisation of this project, especially to AGS, Ferrovial and Ci3 colleagues. 

\section*{Funding}

This work was supported by the EIT Digital under the action A/RPORTWIN of Digital Innovation Factory - Digital Tech with project number 20228, and the Programa Propio UPM.



\bibliographystyle{elsarticle-num} 
\bibliography{biblio.bib}





\end{document}